\documentclass[letter]{aa}  

\usepackage{graphicx}
\usepackage{txfonts}
\usepackage{tabularx}
\usepackage{amsfonts}
\usepackage{bbold}
\usepackage{color}
\usepackage{transparent}
\usepackage{hyperref}
\usepackage{transparent}
\usepackage{rotating}
\usepackage{caption}
\usepackage{subcaption}

\usepackage{graphicx}	
\usepackage{amsmath}	
\usepackage{amssymb}	
\usepackage{tablefootnote}
\usepackage[flushleft]{threeparttable}
\usepackage{dblfloatfix} 

\usepackage{natbib}
\bibpunct{(}{)}{;}{a}{}{,} 

\newcommand{\sgx}{SgXB\xspace}
\newcommand{\ulx}{ULX\xspace}

\newcommand*{\rlof}{RLOF\@\xspace}
\newcommand*{\ns}{NS\@\xspace}
\newcommand*{\bh}{BH\@\xspace}
\newcommand*{\eg}{e.g.\@\xspace}
\newcommand*{\ie}{i.e.\@\xspace}

\newcommand*\diff{\mathop{}\!\mathrm{d}}
\newcommand{\mystar}{{\fontfamily{lmr}\selectfont$\star$}}
\newcommand*{\msun}{$M_{\odot}$\@\xspace}
\newcommand*{\mdotstar}{$\dot{M}_{\text{\mystar}}$\@\xspace}

\newcommand*{\ledd}{$L_{\text{Edd}}$\@\xspace}

\begin{document} 

   \title{Wind Roche lobe overflow in high mass X-ray binaries}

   \subtitle{A possible mass transfer mechanism for Ultraluminous X-ray sources}

   \author{I. El Mellah
          \inst{1}
          \and
          J. O. Sundqvist
          \inst{2}
          \and
          R. Keppens
          \inst{1}
          }

   \institute{Centre for mathematical Plasma Astrophysics, 
   			 Department of Mathematics, KU Leuven, 
   			 Celestijnenlaan 200B, B-3001 Leuven, Belgium\\
              \email{ileyk.elmellah@kuleuven.be}
         \and
             KU Leuven, Instituut voor Sterrenkunde, 
             Celestijnenlaan 200D, B-3001 Leuven, Belgium
             }

   \date{Received ...; accepted ...}

 
  \abstract{Ultra-luminous X-ray sources (\ulx) have so high X-ray luminosities that they were long thought to be accreting intermediate mass black holes. Yet, some \ulx have been shown to display periodic modulations and coherent pulsations, suggestive of a neutron star in orbit around a stellar companion and accreting at super-Eddington rates. In this letter, we propose that the mass transfer in \ulx could be qualitatively the same as in Supergiant X-ray binaries (\sgx), with a wind from the donor star highly beamed towards the compact object. Since the star does not fill its Roche lobe, this mass transfer mechanism known as "wind Roche lobe overflow" can remain stable even for large mass ratios. Based on realistic acceleration profiles derived from spectral observations and modeling of the stellar wind, we compute the bulk motion of the wind to evaluate the fraction of the stellar mass outflow captured by the compact object. We identify the orbital and stellar conditions for a \sgx to transfer mass at rates matching the expectations for \ulx and show that the transition from \sgx to \ulx luminosity levels is progressive. These results indicate that a high stellar Roche lobe filling factor is not necessary to funnel large quantities of material into the Roche lobe of the accretor. Large stellar mass loss rates such as the ones from the Wolf-Rayet star in M101 ULX-1 or the late B9 Supergiant in NGC 7793 P13 are enough to lead to a highly beamed wind and a significantly enhanced mass transfer rate.}


   \maketitle
%

\section{Introduction}

Ultra-luminous X-ray sources are spatially unresolved persistent sources with luminosities in excess of 10$^{39}$erg$\cdot$s$^{-1}$ \citep[for a recent review see][]{Kaaret2017}. This X-ray luminosity threshold corresponds approximately to the Eddington luminosity \ledd of a 10\msun black hole (\bh), the limit above which isotropic accretion onto a body of this mass is thought to be self-regulated by the radiative field it produces \citep{Rappaport2005}. They are found off-centered in galaxies within a couple of 10Mpc, ruling against a supermassive \bh origin. If the emission is not beamed, sub-Eddington accretion can only be sustained for accretors of at least several 10 to several 100\msun, suggestive of the long awaited population of intermediate mass \bh \citep{Colbert1999}. Hints in favor of the existence of intermediate mass \bh recently emerged : the accretor in the ULX M82 X-1 \citep{Brightman2016} has been shown to be a 20 to 35\msun \bh while the observation of gravitational waves emitted by merging compact objects unearthed \bh of several 10\msun \citep{Abbott2016}.

However, the detections of coherent pulsations and cyclotron resonance scattering features from several ULX demonstrate that other ULX host super-Eddington accreting neutron stars \citep[\ns,][]{Bachetti2014,Furst2016,Israel2017,Carpano2018,Brightman2018}. In one of them, NGC 7793 P13 (hereafter P13), \cite{Motch2014} identified a stellar spectrum consistent with a 20\msun B9 Ia star, in a $\sim$64 days orbit with an X-ray source they assumed to be a \bh but which later turned out to be an accreting \ns \citep{Furst2016}. It was further argued that the star has to fill its Roche lobe because the stellar mass loss rate would be too low and wind accretion would not be able to lead to a significant fraction of the stellar wind being captured by the accretor. On the other hand, \cite{Liu2013} showed that in M101 ULX-1 (hereafter M101), the Helium emission lines are best explained by a Wolf-Rayet donor star twice smaller than its Roche lobe. This suggests that mass transfer rates suitable for \ulx X-ray luminosity levels are possible without Roche lobe overflow (\rlof). Finally, in two ULX, near-infrared counterparts consistent with red supergiant donors have been identified \citep{Heida2015,Heida2016}.

Most ULX hosting a stellar mass accretor are now thought to be the high mass accretion rate end of the Supergiant X-ray binaries (\sgx), where the wind from a supergiant donor star acts as a reservoir of matter tapped by the orbiting compact object. The X-ray luminosity functions of \sgx and \ulx follow the same power-law, without apparent break \citep{Gilfanov2004,Swartz2011}. Super-orbital periods are observed in \sgx \citep{Corbet2013} and in \ulx \citep{Walton2016,Fuerst2018}. The main spectral differences between \sgx and \ulx can be attributed either to the nature of the donor or to different accretion geometry in the immediate vicinity of the accretor \citep{Kaaret2017}. All these elements support the idea that both types of objects belong to the same population and that the mass transfer mechanism at the orbital scale might be qualitatively the same.

The final absolute X-ray luminosity $L_X$ released by accretion onto a compact object fed by a stellar companion depends on (i) the stellar mass loss rate \mdotstar, (ii) the rate $\mu$\mdotstar at which mass is transferred from the star into the domain of gravitational influence of the accretor, (iii) the mass which actually ends up being accreted onto the compact object and (iv) the efficiency $\zeta$ of the conversion of accreted mass to radiation, set here to 10$\%$ \citep{Kaaret2017}. The relevance of the classical \ledd limit in \ulx is a matter of debate. For instance, it is derived assuming a smooth stellar wind outflow but since radiation line-driving leads naturally to clumpy and porous winds \citep{Sundqvist2017,Owocki2018}, this may also allow for super-Eddington accretion \citep{Dotan2010}.

In this letter, we set aside the question of how super-Eddington accretion itself proceeds in the vicinity of the accretor \ie how the compact object can accrete at a rate leading to $L_X>$\ledd. Rather, we ask whether stellar material can be transferred to the compact object at a rate $\dot{M}$ high enough to reach the ULX luminosity level, without necessarily assuming \rlof. In the context of symbiotic binaries, \cite{Mohamed2007} showed that a wind speed low enough compared to the orbital speed could lead to a significant enhancement of the mass transfer rate. This mechanism, known as wind Roche lobe overflow (wind-\rlof), is characterized by a strongly beamed wind in the orbital plane and towards the accretor. This mass transfer does not experience runaway \rlof expected for high mass ratios since the process is not conservative and the star does not fill its Roche lobe.

\section{Stellar winds in SgXB}
\label{sec:}


We solve the ballistic equation of motion in the three-dimensional co-rotating frame of a binary system, for test-masses starting at the stellar surface and subjected to the two gravitational forces, the non-inertial forces and the wind acceleration. We use the numerical integrator described in \cite{ElMellah2016a}, with a similar geometry of the problem (see their Figure 2). In \sgx hosting an O/B Supergiant, the wind launching mechanism relies on the resonant absorption of UV photons by partially ionized metal ions \citep{Lucy1970,Castor1975}. In an isotropic situation, this leads to radial velocity profiles which can be well approximated by a $\beta$-law \citep{Puls2008} : 
\begin{equation}
\varv_{\beta}(r)=\varv_{\infty}\left(1-R_{\text{\mystar}}/r\right)^{\beta}
\end{equation}
with $R_{\text{\mystar}}$ the stellar radius, $\varv_{\infty}$ the terminal wind speed and $\beta$ is a positive exponent which represents the efficiency of the acceleration \ie how fast the wind reaches its terminal speed : the lower $\beta$, the earlier $\varv_{\infty}$ is matched. Notice that, although the wind launching mechanism of cool stars is still largely unknown, the wind velocity profiles observed around red supergiants and asymptotic giant branch stars turn out to be in reasonable agreement with this law beyond the condensation radius \citep{Decin2006,Decin2010}. The terminal wind speeds are much lower than for winds from hot stars but they display similar $\beta$ exponents and mass loss rates above 10$^{-5}$\msun$\cdot$yr$^{-1}$ are common \citep{DeBeck2010}. Many uncertainties subsist on the velocity profile of line-driven winds in \sgx, as illustrated by the case of the hot supergiant HD 77581, the donor star in the classic \sgx Vela X-1. From observations of UV spectral lines, \cite{Gimenez-Garcia2016} derived a best $\beta$-law with $\varv_{\infty}\sim 700$km$\cdot$s$^{-1}$ and $\beta=1$ while \cite{Sander2017} computed the hydrodynamic atmosphere solution for the wind stratification, best-fitted by a $\beta$-law with $\varv_{\infty}\sim 500$km$\cdot$s$^{-1}$ and $\beta\sim 2.3$. Given these uncertainties, we evaluate the wind acceleration in an empirical manner by relying on representative velocity profiles from which we can derive steady radial line-driven accelerations. We also need to assume that the departure from the isotropic case due to the presence of an orbiting accretor does not significantly alter the radial component of the wind acceleration, still given by $\varv_{\beta}\diff_r\varv_{\beta}$. 

To evaluate the fraction of wind captured $\mu=\dot{M}/$\mdotstar, we need to define the region of gravitational influence of the accretor. It is given by the Roche lobe if the accretion radius $R_{acc}$ is larger than the Roche lobe radius, and the sphere of radius $R_{acc}$ centered on the accretor otherwise. The accretion radius is an accurate estimate of the effective cross section of a point mass accreting from a planar uniform wind for Mach numbers of the incoming flow above 3 \citep{ElMellah2015}. It reads :
\begin{equation}
R_{acc}=2GM_{\bullet}/v^2_{\beta}(r=a)
\end{equation}
with $a$ the orbital separation, $G$ the constant of gravity and $M_{\bullet}$ the mass of the accretor. In \sgx and \ulx, the accreting compact object lies deep in the wind, in a region where the wind is still accelerating. Given the uncertainties which remain on this regime and the important dependency of the accretion radius on the wind speed, we need to consider a variety of $\beta$ exponents : for instance, in Vela X-1, the uncertainty on the wind velocity profile leads to a discrepancy of $\sim$10 on the accretion radius of the \ns.

In dimensionless form, the solutions of the equation of motion depend only on the mass ratio $q$, the filling factor $f$ \citep[the ratio of the stellar radius by the Roche lobe radius,][]{Eggleton1983}, the $\beta$ exponent and the ratio of the terminal wind speed by the orbital speed $\eta=\varv_{\infty}/\varv_{orb}$, with $\varv_{orb}=2\pi a/P_{\text{orb}}$ and $P_{\text{orb}}$ the orbital period. In particular, these 4 parameters entirely determine the ratio $\mu$ of stellar wind significantly beamed towards the accretor. We quantify $\mu$ by monitoring the fraction of integrated streamlines entering the domain of gravitational influence of the accretor.


%
%
%

\begin{figure*}[!t]
\centering
\includegraphics[width=2\columnwidth]{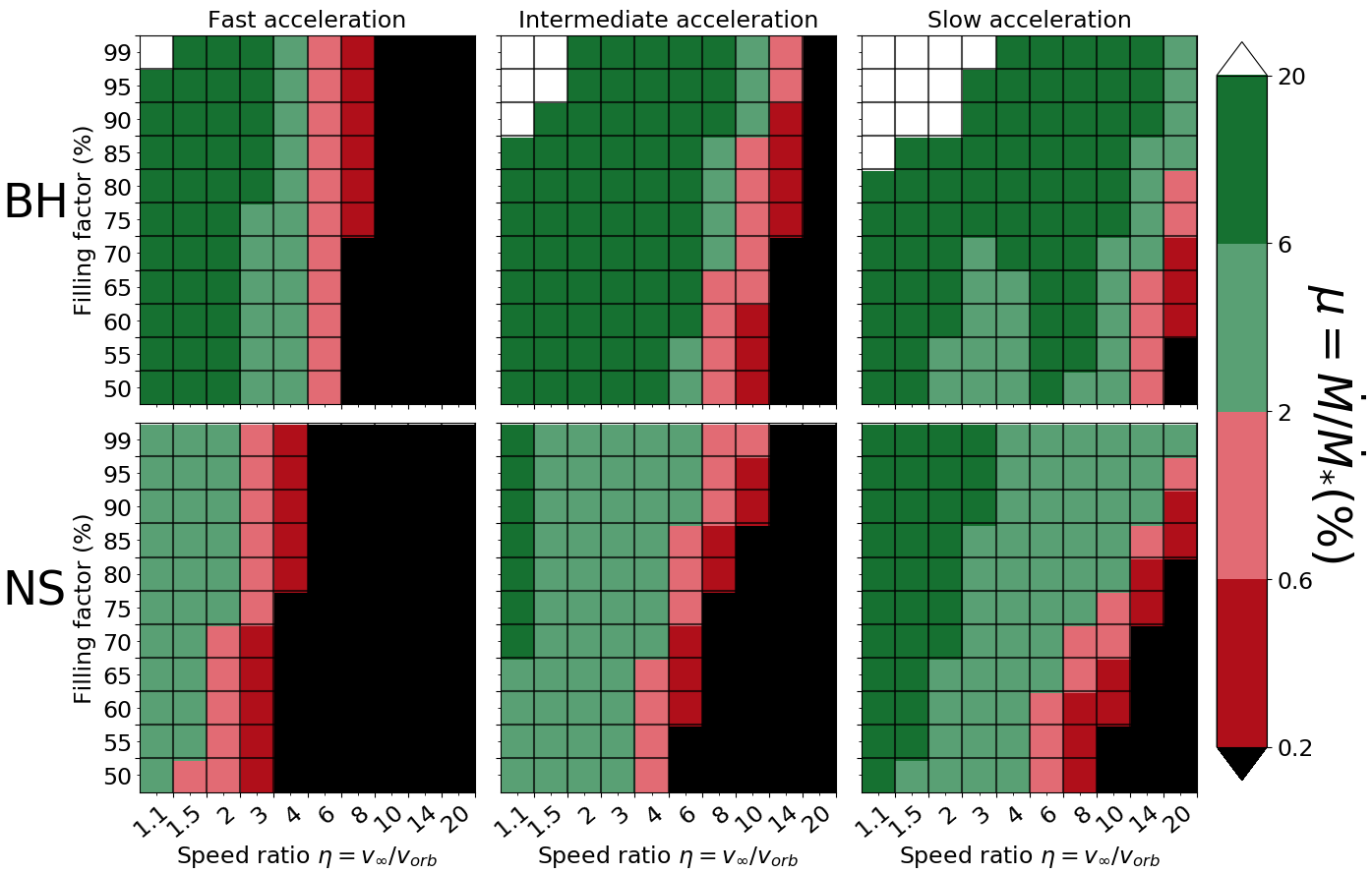}
\caption{Logarithmic color maps of the fraction $\mu$ of stellar wind captured by the accretor as a function of the stellar filling factor and of the ratio of the terminal wind speed by the orbital speed. From left to right, $\beta=$1, 2 and 3. The upper (resp. lower) row stands for a mass ratio of 2 (resp. 15) typical of a \bh (resp. a \ns) accreting from a supergiant companion.}
\label{fig:mdot}
\end{figure*} 

\begin{figure}
\centering
\includegraphics[width=1\columnwidth]{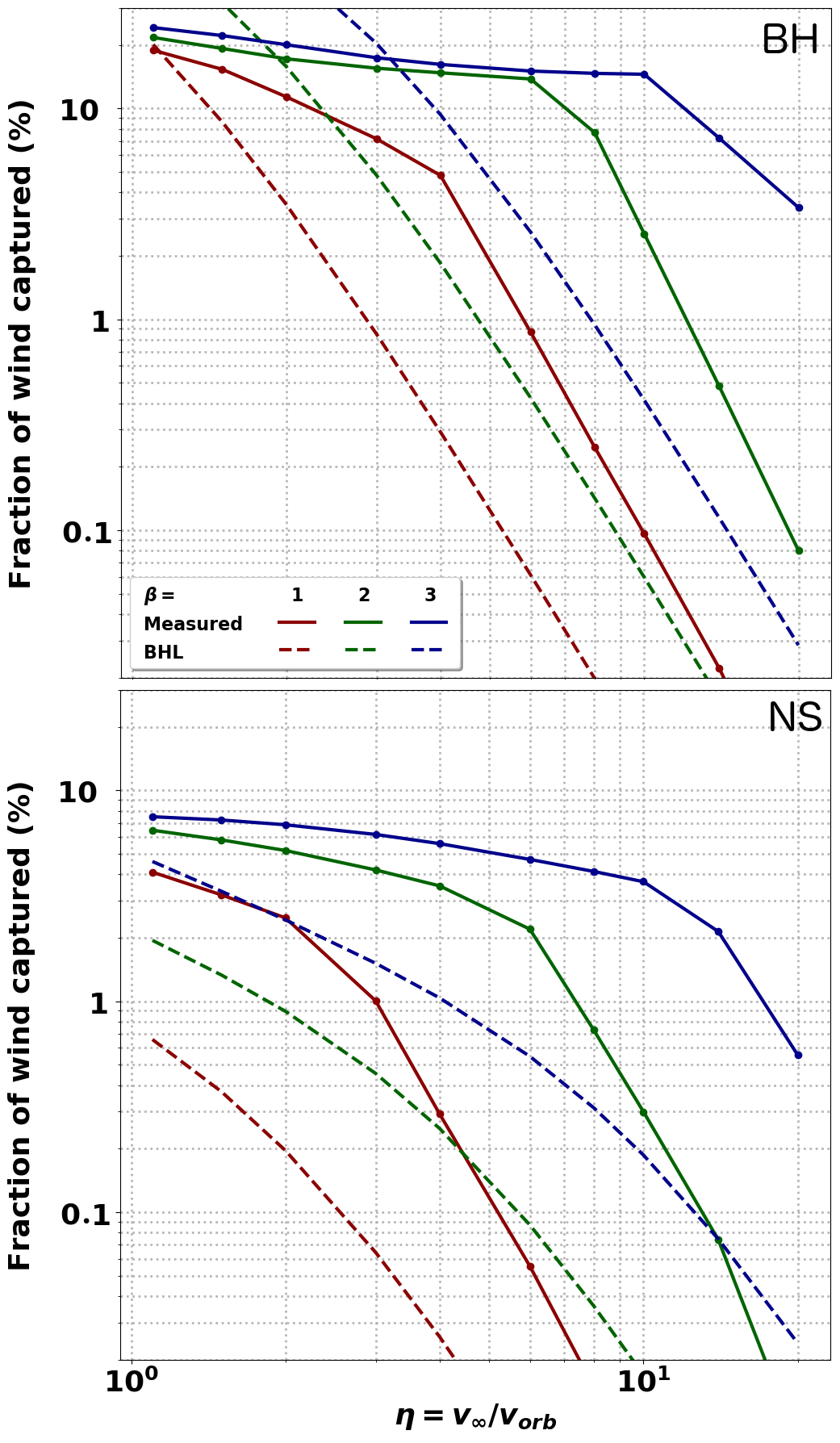}
\caption{Comparison of the mass transfer rates we measured as a function of $\eta$ (solid lines) to the BHL mass accretion rate formula (dashed lines) for a filling factor of 90\% and a mass ratio of 2 (upper panel) and 15 (lower panel). The BHL prescription generally underestimates the mass accretion rate, sometimes by more than an order of magnitude.}
\label{fig:mdot2}
\end{figure} 

\section{Mass transfer via wind-RLOF}
\label{sec:}

\subsection{Fraction of stellar wind available for accretion}
\label{sec:}

We probe a range of parameters realistic for \ulx, whether the donor star is a Wolf-Rayet or a supergiant. The few orbital periods and mass measures available indicate that the orbital speed might vary by a factor of a few from one \ulx to another, while the terminal wind speeds of cool supergiants are generally more than an order of magnitude lower than for hot supergiants and Wolf-Rayet stars : here, we span values of $\eta$ from 1.1 to 20. We work with a $\beta$ exponent of 1, 2 and 3, with 3 giving the most progressive velocity increase to the terminal speed. Observed filling factors lie between 50\% and 100\% (\rlof). Two extreme mass ratios have been considered, $q=$2 and $q=$15, which correspond respectively to a stellar mass \bh and a \ns accreting from a $\sim$20\msun supergiant.

Figure\,\ref{fig:mdot} illustrates the evolution of the fraction of wind captured with the 4 dimensionless parameters of the problem. The main parameter is $\eta$ : when the terminal wind speed gets a few times larger than the orbital speed, the effective cross-section set by the accretion radius decreases quickly and much below the radius of the Roche lobe of the accretor. This effect is tempered for slowly accelerating winds (\ie for high $\beta$) which, provided the distance from the photosphere to the accretor is small (\ie for large $f$), are still far from their terminal wind speeds when they enter the Roche lobe of the accretor. On the contrary, the dependency of $\mu$ on the filling factor vanishes when the wind accelerates quickly because then, by the time it reaches the accretor, the wind has almost reached its terminal speed. Slow wind launching leads to a significant contribution of the high latitude stellar regions : the flow is strongly beamed in the orbital plane and the value of $\mu$ surges. 

We now compare our results to the commonly used Bondi-Hoyle-Lyttleton (BHL) mass accretion rate formula \citep{Bondi1944,Edgar:2004ip} :
\begin{equation}
\dot{M}_{BHL}=\pi R_{BHL}^2 \varv_{rel} \rho_{\bullet} 
\label{eq:BHL}
\end{equation}
where $\dot{M}_{BHL}$ is the BHL mass accretion rate, $\varv_{rel}=\sqrt{\varv_{\beta}^2+\varv_{\bullet}^2}$ is the relative speed between the wind and the compact object whose orbital speed is given by $\varv_{\bullet}=\varv_{orb}\left[q/\left(1+q\right)\right]$ and $R_{BHL}=2GM_{\bullet}/\varv_{rel}^2$ is the modified accretion radius. In the literature, the density of the wind at orbital separation $\rho_{\bullet}$ is evaluated assuming an isotropic dilution of the wind around the star. We can now write the fraction of wind captured according to the BHL formula \eqref{eq:BHL}, $\mu_{BHL}=\dot{M}_{BHL}/\dot{M}_{\text{\mystar}}$, as a function of the 4 dimensionless parameters :
\begin{equation}
\mu_{BHL}=\frac{\left(1+q\right)/q^3}{\eta\left(1-f\mathcal{E}\right)^{\beta}\left[1+\left(\eta\left(1+q\right)\left(1-f\mathcal{E}\right)^{\beta}/q\right)\right]^{3/2}}
\end{equation}
where $\mathcal{E}$ is the ratio of the stellar Roche lobe radius by the orbital separation given by \cite{Eggleton1983} and function only of $q$. In Figure\,\ref{fig:mdot2}, we plotted with dashed lines the evolution of $\mu_{BHL}$ with the speed ratio $\eta$ for different mass ratios and $\beta$ exponents and compared it with our results. Since our setup does account for the geometrical beaming of matter, we find values of $\mu$ systematically higher, except in when the wind speed at the orbital separation is so low compared to the orbital speed than the effective cross section given by the BHL formula is larger than the orbital separation and loose any meaning. 



\subsection{Accretion luminosity}
\label{sec:}

Another insight revealed by this analysis is the identification of the configurations where \ulx can not be the product of mass transfer via wind-\rlof and can be explained only if the star fills its Roche lobe. In Figure\,\ref{fig:mdot}, we represented in black the regions where $\mu$ is inferior to 0.2\%, the minimum fraction of stellar wind captured under which mass transfer without \rlof is not able to produce an X-ray luminosity of 10$^{39}$erg$\cdot$s$^{-1}$, even for a stellar mass loss rate as high as 10$^{-4}$\msun$\cdot$yr$^{-1}$ :
\begin{equation}
\mu_{\text{min}}\sim 0.2\%\left(\frac{10\%}{\eta}\right)\left(\frac{10^{-4}\text{\msun}\cdot\text{yr}^{-1}}{\text{\mdotstar}}\right)\left(\frac{L_X}{10^{39}erg\cdot s^{-1}}\right)
\label{eq:lim}
\end{equation} 
Given the maximal $\mu$ we found at $\sim$20\%, this analysis suggests that mass transfer via wind-\rlof in a \ulx is only possible for stellar mass loss rates above $\sim$10$^{-6}$\msun$\cdot$yr$^{-1}$.

Rather than covering the large scope of regimes possible, let us focus on 3 donor stars in one \sgx and two \ulx. In the \sgx Vela X-1, \cite{Gimenez-Garcia2016} reported a stellar mass loss rate $\dot{M}_{\text{\mystar}}\sim6.3\cdot 10^{-7}$\msun$\cdot$yr$^{-1}$ for the B0.5 Ib supergiant donor star. In M101, \cite{Liu2013} inferred a mass loss rate of 2$\cdot$10$^{-5}$\msun$\cdot$yr$^{-1}$ for the Wolf-Rayet donor. In P13, the stellar mass loss rate of the B9 Ia supergiant donor star might well have been underestimated : in \cite{Motch2014}, it is assumed that the donor star fills its Roche lobe on the basis that the wind mass loss rate derived by \cite{Kudritzki1999} would be inferior to 10$^{-6}$\msun$\cdot$yr$^{-1}$. However, \cite{Kudritzki1999} did not include late type B supergiants in their study because of the difficulty to measure their terminal wind speed \citep{Howarth1997}. \cite{Vink2000a} and \cite{Vink2001} provided fits for the mass loss rate and the terminal wind speed which extend below the second bi-stability jump \citep[\ie for effective temperatures $T_{eff}\lesssim12kK$,][]{Lamers2000} : with the values reported by \cite{Motch2014} ($T_{eff}\sim$11kK, stellar luminosity L$_{\text{\mystar}}\sim10^{5.1}L_{\odot}$ and stellar mass $\sim$20\msun), the mass loss rate is $\sim$10$^{-5}$\msun$\cdot$yr$^{-1}$ for solar metallicity, comparable to the one found in M101.

Let us now estimate the 4 dimensionless parameters for each system. For P13, \cite{Fuerst2018} determined orbital elements which indicate, in combination with the stellar parameters above, $f>90\%$, $q\sim 15$ and $\varv_{orb}=$100 to 200km$\cdot$s$^{-1}$ (for an inclination angle of respectively the maximum one set by the absence of eclipse, $\sim$55$^{\circ}$, and 20$^{\circ}$). At the bi-stability jump, the terminal wind speed should be between 0.7 and 1.3 times the escape velocity at the stellar surface \ie $\varv_{\infty}=$150 to 300km$\cdot$s$^{-1}$, hence $\eta=1$ to 3. According to Figure\,\ref{fig:mdot}, $\mu_{\text{min}}=6\%$ for the observed $L_X\sim 3\cdot 10^{39}$erg$\cdot$s$^{-1}$ is reached only if $\beta\geqslant 2$ and for minimal speed ratios $\eta\leqslant 1.5$. Stellar material at a sufficiently high rate could then be supplied via wind-\rlof, although marginally. 

In M101, the mass function indicates that the accretor must be a \bh, with $q\sim 2$ for an intermediate orbital inclination \citep{Liu2013}. Their best fit to explain the Helium emission lines leads to a filling factor of at most $f=50\%$ and a terminal wind speed of 1,300 km$\cdot$s$^{-1}$, 3 to 4 times larger than the orbital speed. It leads to $\mu\gtrsim 6\%$, in agreement with the mass accretion rate required to reproduce the \ulx luminosity with the stellar mass loss rate observed. Since the star does not fill its Roche lobe at all, this \ulx can only be explained with the wind-\rlof mass transfer mechanism.

Vela X-1 has dimensionless parameters which lie in the same ranges as P13, but is crippled by its comparatively low stellar mass loss rate and possible effects which prevents a significant fraction of the mass captured from eventually being accreted. For instance, hydrodynamics simulations within the Roche lobe of the compact object suggest that, when the flow can not radiate away the energy it gains as it is adiabatically compressed and/or shocked, the effective mass accretion rate can be lowered by a factor $\sim$10 compared to the mass capture rate $\mu$\mdotstar computed here \citep{ElMellah2018}. Since the wind speeds are comparable in \ulx and \sgx, the flow density in Vela X-1 must be much lower than in P13 or M101, leading to a less efficient radiative cooling. Also, the very high X-ray luminosities in \ulx are expected to considerably ionize the wind, even for orbital separations as large as in P13 \citep{Blondin1991,Manousakis2015c}. At high ionization, the wind acceleration is inhibited which participates in the efficient accretion of the material entering the sphere of gravitational influence of the accretor. Interestingly enough, \cite{Ho1987} already pinpointed that a high luminosity solution could exist when the wind was highly ionized. For Vela X-1, the low luminosity solution matches the few 10$^{36}$erg$\cdot$s$^{-1}$ observed but once the wind is in the high ionization stage, the X-ray luminosity is 2,000 times larger, matching a \ulx level, although such a flare has never been observed in Vela X-1. To a lesser extent, the fraction of mass captured ending up being accreted can also be altered by the clumpiness of the wind \citep{Sundqvist2017,ElMellah}. These combined effects might explain why many \sgx such as Vela X-1 have luminosities much lower than the Eddington luminosity of a \ns or a stellar mass \bh, in spite of similar dimensionless parameters.


\section{Discussion and conclusions}
\label{sec:}


We showed that efficient mass transfer without \rlof could occur in a binary system from a Supergiant donor star to its compact accretor in \sgx. When the wind is slow enough compared to the orbital speed to see its dynamics significantly altered by the Roche potential, it is beamed towards the accretor. Under these conditions, the wind density in the orbital plane is enhanced, which considerably increases the fraction of the stellar wind captured up to several 10\%. If the mass loss rate of the star is large enough, it can result in a mass available for accretion being supplied at a rate suitable to produce X-ray accretion luminosity of the order of the ones observed in \ulx. The final mass accretion rate though depends also on the compressibility of the flow, on the X-ray ionizing feedback and on the clumpiness of the wind being captured.

Some \ulx are also expected to undergo \rlof mass transfer which is bound to be unstable when the mass ratio is as high as in P13 \citep{King2002,Rappaport2005}. This mechanism is the most reliable scenario when the stellar mass loss rate is found to be smaller than 10$^{-6}$\msun$\cdot$yr$^{-1}$ or for hyperluminous X-ray sources \citep{Webb2017}. Alternatively, in \ulx where the donor star has a sufficient mass loss rate and is found to not fill its Roche lobe such as M101 ULX-1, wind-\rlof provides a good description of the geometry of the flow. Regardless of the nature of the wind-launching mechanism, the same arguments apply to the red supergiant donors identified in a couple of ULX \citep{Heida2015,Heida2016}. Due to their low terminal speeds and high mass loss rates, they are good candidates to transfer large amounts of mass via wind-\rlof \citep{Mohamed2007,DeVal-Borro2017} and to be well described by the formalism developed here, even for long orbital periods.

With this letter, we emphasize on the compelling need to improve our knowledge of the nature of the donor star and in particular of its mass loss to understand not only \sgx but also the high mass accretion tail of the distribution, \ulx. In this attempt, observational campaigns have been carried out \cite[see \eg][]{Heida2014}. Theoretical arguments have been made to constrain the donor star but they often presuppose a \rlof mass transfer and, thus, might not hold in general for \ulx \citep{Karino2017}. If the stellar mass loss rate is too low or the source too bright (typically, hyperluminous X-ray sources), wind-\rlof can not channel enough matter but otherwise, when the wind is slow enough, it is a fully viable scenario to enhance the mass transfer rate. 

In several \ulx, the presence of a disc has been inferred from super-orbital modulations or systematic spinning up of the \ns \citep[\eg in P13,][]{Fuerst2018}. A disc can form only if the wind is slow enough \citep{Illarionov1975} but in a pure wind accretion situation, the mass accretion rate increases quickly when the wind speed decreases. \cite{Tutukov2016} noticed that in IC 10 X-1 and NGC 300 X-1, two systems where a stellar mass \bh accretes from a Wolf-Rayet companion, no wind speed could explain both the accretion disc and the limited X-ray luminosity of 10$^{38}$erg$\cdot$s$^{-1}$. The wind-\rlof mechanism, fully compatible with the formation of a wind captured disc while not necessarily leading to large mass accretion rates, solves this issue \citep{ElMellah2018}. 

Last, due to its enhanced efficacity while still being stable, this mass transfer mechanism could have important consequences on population synthesis of binary systems. The results presented here provide the community with estimates for the mass transfer rate as a function of a limited number of dimensionless parameters.




\begin{acknowledgements}
IEM is grateful to Marianne Heida for insightful exchanges on the possible scenarios leading to ultra-luminous X-ray sources. IEM wishes to thank Andreas Sander and Jorick Vink for their estimates on the stellar mass loss rates of hot stars. IEM has received funding from the Research Foundation Flanders (FWO) and the European Union's Horizon 2020 research and innovation program under the Marie Sk\l odowska-Curie grant agreement No 665501. IEM and JOS thank the Instituto de F\'{i}sica de Cantabria for its hospitality and for sponsoring a meeting which brought together the massive stars and X-ray binaries communities.
\end{acknowledgements}


\bibliographystyle{aa} 
\begin{tiny}
\bibliography{/Users/Ileyk/Documents/Bibtex/article_ULX_no_url}
\end{tiny}

\end{document}